\documentclass[12pt,letterpaper]{article}
\usepackage{epsfig,rotating,setspace,latexsym,amsmath,epsf,amssymb,bm,theorem}
\usepackage{cite,appendix}

\usepackage{amsfonts}

\title{Capacity-Equivocation Region of the Gaussian MIMO Wiretap Channel\thanks{This work was supported by NSF Grants CCF
04-47613, CCF 05-14846, CNS 07-16311 and CCF 07-29127.}}

\author{Ersen Ekrem \qquad Sennur Ulukus \\
\normalsize Department of Electrical and Computer Engineering\\
\normalsize University of Maryland, College Park, MD 20742 \\
\normalsize {\it ersen@umd.edu} \qquad {\it ulukus@umd.edu}}

\newcommand{\bblambda}{\bm \Lambda}

\newcommand{\bbdelta}{\bm \Delta}
\newcommand{\bbsigma}{\bm \Sigma}

\newcommand{\bbi}{{\mathbf{I}}}
\newcommand{\bzero}{{\mathbf{0}}}
\newcommand{\bbv}{{\mathbf{V}}}

\newcommand{\bbh}{{\mathbf{H}}}

\newcommand{\bbm}{{\mathbf{M}}}

\newcommand{\bbk}{{\mathbf{K}}}
\newcommand{\bbz}{{\mathbf{Z}}}
\newcommand{\bbn}{{\mathbf{N}}}

\newcommand{\bba}{{\mathbf{A}}}

\newcommand{\bbt}{{\mathbf{T}}}
\newcommand{\bbb}{{\mathbf{B}}}

\newcommand{\bbs}{{\mathbf{S}}}

\newcommand{\bbu}{{\mathbf{U}}}

\newcommand{\bbx}{{\mathbf{X}}}

\newcommand{\bby}{{\mathbf{Y}}}

\newtheorem{Theo}{Theorem}

\newtheorem{Lem}{Lemma}
\newtheorem{Cor}{Corollary}

\begin{document}

%\IEEEoverridecommandlockouts

\maketitle

\begin{abstract}
We study the Gaussian multiple-input multiple-output (MIMO)
wiretap channel, which consists of a transmitter, a legitimate
user, and an eavesdropper. In this channel, the transmitter sends
a common message to both the legitimate user and the eavesdropper.
In addition to this common message, the legitimate user receives a
private message, which is desired to be kept hidden as much as
possible from the eavesdropper. We obtain the entire
capacity-equivocation region of the Gaussian MIMO wiretap channel.
This region contains all achievable common message, private
message, and private message's equivocation (secrecy) rates. In
particular, we show the sufficiency of jointly Gaussian auxiliary
random variables and channel input to evaluate the existing
single-letter description of the capacity-equivocation region due
to Csiszar-Korner.

\end{abstract}

\newpage

\section{Introduction}
We consider the Gaussian multiple-input multiple-output (MIMO)
wiretap channel, which consists of a transmitter, a legitimate
user, and an eavesdropper. In this channel, the transmitter sends
a common message to both the legitimate user and the eavesdropper
in addition to a private message which is directed to only the
legitimate user. There is a secrecy concern regarding this private
message in the sense that the private message needs to be kept
secret as much as possible from the eavesdropper. The secrecy of
the private message is measured by its equivocation at the
eavesdropper.

Here, we obtain the capacity-equivocation region of the Gaussian
MIMO wiretap channel. This region contains all achievable rate
triples $(R_0,R_1,R_e)$, where $R_0$ denotes the common message
rate, $R_1$ denotes the private message rate, and $R_e$ denotes
the private message's equivocation (secrecy) rate. In fact, this
region is known in a single-letter form due to~\cite{Korner}. In
this work, we show that jointly Gaussian auxiliary random
variables and channel input are sufficient to evaluate this
single-letter description for the capacity-equivocation region of
the Gaussian MIMO wiretap channel. We prove the sufficiency of the
jointly Gaussian auxiliary random variables and channel input by
using channel enhancement~\cite{Shamai_MIMO} and an extremal
inequality from~\cite{Liu_Compound}. In our proof, we also use the
equivalence between the Gaussian MIMO wiretap channel and the
Gaussian MIMO wiretap channel with {\em public}
messages~\cite[Problem~33-c]{Csiszar_book},~\cite{Ruoheng_Equivocation}.
In the latter channel model, the transmitter has three messages, a
common, a confidential, and a public message. The common message
is sent to both the legitimate user and the eavesdropper, while
the confidential and public messages are directed to only the
legitimate user. Here, the confidential message needs to be
transmitted in perfect secrecy, whereas there is no secrecy
constraint on the public message. Since the Gaussian MIMO wiretap
channel and the Gaussian MIMO wiretap channel with public messages
are equivalent, i.e., there is a one-to-one correspondence between
the capacity regions of these two models, in our proof, we obtain
the capacity region of the Gaussian MIMO wiretap channel with
public messages, which, in turn, gives us the
capacity-equivocation region of the Gaussian MIMO wiretap channel.

Our result subsumes the following previous findings about the
capacity-equivocation region of the Gaussian MIMO wiretap channel:
i) The secrecy capacity of this channel, i.e., $\max R_1$ when
$R_0=0,R_e=R_1$, is obtained in \cite{Hassibi,Wornell} for the
general case, and in \cite{Ulukus} for the 2-2-1 case. ii) The
common and confidential rate region under perfect secrecy, i.e.,
$(R_0,R_1)$ region with $R_e=R_1$, is obtained
in~\cite{Liu_Common_Confidential}. iii) The capacity-equivocation
region without a common message, i.e., $(R_1,R_e)$ region with
$R_0=0$, is obtained in~\cite{Ruoheng_Equivocation}. iv) The
capacity region of the Gaussian MIMO broadcast channel with
degraded message sets without a secrecy concern, i.e., $(R_0,R_1)$
region with no consideration on $R_e$, is obtained
in~\cite{Hannan_Common}. Here, we obtain the entire
$(R_0,R_1,R_e)$ region.

\section{Discrete Memoryless Wiretap Channels}
The discrete memoryless wiretap channel consists of a transmitter,
a legitimate user and an eavesdropper. The channel transition
probability is denoted by $p(y,z|x)$, where $x\in\mathcal{X}$ is
the channel input, $y\in\mathcal{Y}$ is the legitimate user's
observation, and $z\in\mathcal{Z}$ is the eavesdropper's
observation. We consider the following scenario for the discrete
memoryless wiretap channel: The transmitter sends a common message
to both the legitimate user and the eavesdropper, and a private
message to the legitimate user which is desired to be kept hidden
as much as possible from the eavesdropper.

An $(n,2^{nR_0},2^{nR_1})$ code for this channel consists of two
message sets $\mathcal{W}_0=\{1,\ldots,2^{nR_0}\}$,
$\mathcal{W}_1=\{1,\ldots,2^{nR_1}\}$, one encoder at the
transmitter $f:\mathcal{W}_0\times \mathcal{W}_1\rightarrow
\mathcal{X}^n$, one decoder at the legitimate user
$g_u:\mathcal{Y}^n\rightarrow \mathcal{W}_0\times\mathcal{W}_1$,
and one decoder at the eavesdropper $g_e:\mathcal{Z}^n\rightarrow
W_0$. The probability of error is defined as
$P_e^n=\max\{P_{e,u}^n,P_{e,e}^n\}$, where
$P_{e,u}^n=\Pr[g_u(Y^n)\neq (W_0,W_1)], P_{e,e}^n=\Pr[g_e(Z^n)\neq
W_0]$, and $W_j$ is a uniformly distributed random variable in
$\mathcal{W}_j,~j=0,1$. The secrecy of the legitimate user's
private message is measured by its equivocation at the
eavesdropper~\cite{Korner,Wyner}, i.e.,
\begin{align}
\frac{1}{n}H(W_1|W_0,Z^n)
\end{align}
A rate triple $(R_0,R_1,R_e)$ is said to be achievable if there
exists an $(n,2^{nR_0},2^{nR_1})$ code such that
$\lim_{n\rightarrow \infty}P_e^n=0$, and
\begin{align}
R_e=\lim_{n\rightarrow \infty}\frac{1}{n}H(W_1|W_0,Z^n)
\end{align}
The capacity-equivocation region of the discrete memoryless
wiretap channel is defined as the convex closure of all achievable
rate triples $(R_0,R_1,R_e)$, and denoted by $\mathcal{C}$. The
capacity-equivocation region of the discrete memoryless wiretap
channel, which is obtained in~\cite{Korner}, is stated in the
following theorem.
\begin{Theo}{\bf(\!\cite[Theorem~1]{Korner})}
\label{theorem_csiszar} The capacity-equivocation region of the
discrete memoryless wiretap channel $\mathcal{C}$ is given by the
union of rate triples $(R_0,R_1,R_e)$ satisfying
\begin{align}
0&\leq R_e\leq R_1 \label{cap_equi_discrete_1}\\
R_e&\leq I(V;Y|U)-I(V;Z|U)\label{cap_equi_discrete_2}\\
R_0+R_1&\leq I(V;Y|U)+\min\{I(U;Y),I(U;Z)\}\label{cap_equi_discrete_3}\\
R_0&\leq \min\{I(U;Y),I(U;Z)\} \label{cap_equi_discrete_4}
\end{align}
for some $U,V,X$ such that
\begin{align}
U\rightarrow V\rightarrow X\rightarrow (Y,Z)
\end{align}
\end{Theo}

We next provide an alternative description for $\mathcal{C}$. This
alternative description will arise as the capacity region of a
different, however related, communication scenario for the
discrete memoryless wiretap channel. In this communication
scenario, the transmitter has three messages, $W_0,W_p,W_s$, where
$W_0$ is sent to both the legitimate user and the eavesdropper,
and $W_s,W_p$ are sent only to the legitimate user. In this
scenario, $W_s$ needs to be transmitted in perfect secrecy, i.e.,
it needs to satisfy
\begin{align}
\lim_{n\rightarrow
\infty}\frac{1}{n}I(W_s;Z^n,W_0)=0\label{perfect_secrecy}
\end{align}
and there is no secrecy constraint on the \emph{public} message
$W_p$. To distinguish this communication scenario from the
previous one, we call the channel model arising from this scenario
the discrete memoryless wiretap channel with \emph{public}
messages. We note that this alternative description for wiretap
channels has been previously considered
in~\cite[Problem~33-c]{Csiszar_book},~\cite{Ruoheng_Equivocation}.

An $(n,2^{nR_0},2^{nR_p},2^{nR_s})$ code for this scenario
consists of three message sets $\mathcal{W}_0=\{1,\ldots,\break
2^{nR_0}\},\mathcal{W}_p=\{1,\ldots,2^{nR_p}\},\mathcal{W}_s=\{1,\ldots,2^{nR_s}\}$,
one encoder at the transmitter
$f:\mathcal{W}_0\times\mathcal{W}_p\times \mathcal{W}_s\rightarrow
\mathcal{X}^n$, one decoder at the legitimate user
$g_u:\mathcal{Y}^n\rightarrow \mathcal{W}_0\times \mathcal{W}_p
\times \mathcal{W}_s$, and one decoder at the eavesdropper
$g_e:\mathcal{Z}^n\rightarrow \mathcal{W}_0$. The probability of
error is defined as $P_e^n=\max\{P_{e,u}^n,P_{e,e}^n\}$, where
$P_{e,u}^n=\Pr[g_{u}(Y^n)\neq (W_0,W_p,W_s)]$ and
$P_{e,e}^n=\Pr[g_{e}(Z^n)\neq W_0]$. A rate triple
$(R_{0},R_p,R_s)$ is said to be achievable if there exists an
$(n,2^{nR_0},2^{nR_p},2^{nR_s})$ code such that
$\lim_{n\rightarrow \infty}P_e^n=0$ and (\ref{perfect_secrecy}) is
satisfied. The capacity region $\mathcal{C}_p$ of the discrete
memoryless wiretap channel with \emph{public} messages is defined
as the convex closure of all achievable rate triples
$(R_0,R_p,R_s)$. The following lemma establishes the equivalence
between $\mathcal{C}$ and $\mathcal{C}_p$.
\begin{Lem}
\label{lemma_equivalence} $(R_0,R_p,R_s)\in \mathcal{C}_p$ iff
$(R_0,R_s+R_p,R_s)\in \mathcal{C}$.
\end{Lem}
The proof of this lemma is given in
Appendix~\ref{proof_of_lemma_equivalence}. Using
Lemma~\ref{lemma_equivalence} and Theorem~\ref{theorem_csiszar},
we can express $\mathcal{C}_p$ as stated in the following theorem.
\begin{Theo}
\label{theorem_csiszar_alternative} The capacity region of the
discrete memoryless wiretap channel with public messages
$\mathcal{C}_p$ is given by the union of rate triples
$(R_0,R_p,R_s)$ satisfying
\begin{align}
0&\leq R_s \leq I(V;Y|U)-I(V;Z|U)\\
R_0+R_p+R_s&\leq I(V;Y|U)+\min\{I(U;Y),I(U;Z)\}\\
R_0&\leq \min\{I(U;Y),I(U;Z)\}
\end{align}
for some $(U,V,X)$ such that
\begin{align}
U\rightarrow V\rightarrow X\rightarrow (Y,Z)
\end{align}
\end{Theo}

\section{Gaussian MIMO Wiretap Channel}
The Gaussian MIMO wiretap channel is defined by
\begin{align}
\bby&=\bbh_Y \bbx+\bbn_Y \label{general_gaussian_mimo_1}\\
\bbz&=\bbh_Z \bbx+\bbn_Z \label{general_gaussian_mimo_2}
\end{align}
where the channel input $\bbx$ is a $t\times 1$ vector, $\bby$ is
an $r_Y\times 1$ column vector denoting the legitimate user's
observation, $\bbz$ is an $r_Z\times 1$ column vector denoting the
eavesdropper's observation, $\bbh_Y,\bbh_Z$ are the channel gain
matrices of sizes $r_Y\times t,r_Z\times t$, respectively, and
$\bbn_Y,\bbn_Z$ are Gaussian random vectors with covariance
matrices $\bbsigma_Y,\bbsigma_Z$\footnote{Without loss of
generality, we can set $\bbsigma_Y=\bbsigma_Z=\bbi$. However, we
let $\bbsigma_Y,\bbsigma_Z$ be arbitrary for ease of
presentation.}, respectively, which are assumed to be strictly
positive-definite, i.e., $\bbsigma_Y\succ
\bzero,\bbsigma_Z\succ\bzero$. We consider a covariance constraint
on the channel input as follows
\begin{align}
E\left[\bbx\bbx^{\top}\right] \preceq \bbs
\label{covariance_constraint}
\end{align}
where $\bbs \succeq \bzero$. The capacity-equivocation region of
the Gaussian MIMO wiretap channel is denoted by
$\mathcal{C}(\bbs)$, and is stated in the following theorem.
\begin{Theo}
\label{theorem_main_result} The capacity-equivocation region of
the Gaussian MIMO wiretap channel $\mathcal{C}(\bbs)$ is given by
the union of rate triples $(R_0,R_1,R_e)$ satisfying
\begin{align}
0\leq R_e&\leq
\frac{1}{2}\log\frac{|\bbh_Y\bbk\bbh_Y^\top+\bbsigma_Y|}{|\bbsigma_Y|}-\frac{1}{2}\log\frac{|\bbh_Z\bbk\bbh_Z^\top+\bbsigma_Z|}{|\bbsigma_Z|}
\\
R_0+R_1&\leq
\frac{1}{2}\log\frac{|\bbh_Y\bbk\bbh_Y^\top+\bbsigma_Y|}{|\bbsigma_Y|}
\nonumber\\
&\quad
+\min\left\{\frac{1}{2}\log\frac{|\bbh_Y\bbs\bbh_Y^\top+\bbsigma_Y|}{|\bbh_Y\bbk\bbh_Y^\top+\bbsigma_Y|},\frac{1}{2}\log\frac{|\bbh_Z\bbs\bbh_Z^\top+\bbsigma_Z|}{|\bbh_Z\bbk\bbh_Z^\top+\bbsigma_Z|}\right\}
\\
R_0&\leq
\min\left\{\frac{1}{2}\log\frac{|\bbh_Y\bbs\bbh_Y^\top+\bbsigma_Y|}{|\bbh_Y\bbk\bbh_Y^\top+\bbsigma_Y|},\frac{1}{2}\log\frac{|\bbh_Z\bbs\bbh_Z^\top+\bbsigma_Z|}{|\bbh_Z\bbk\bbh_Z^\top+\bbsigma_Z|}\right\}
\end{align}
for some positive semi-definite matrix $\bbk$ such that
$\bzero\preceq \bbk\preceq \bbs$.
\end{Theo}
Similar to what we did in the previous section, we can establish
an alternative statement for Theorem~\ref{theorem_main_result} by
considering the Gaussian MIMO wiretap channel with \emph{public}
messages, where the legitimate user's private message is divided
into two parts such that one part (confidential message) needs to
be transmitted in perfect secrecy and there is no secrecy
constraint on the other part (public message). The capacity region
for this alternative scenario is denoted by $\mathcal{C}_p(\bbs)$,
and can be obtained by using Lemma~\ref{lemma_equivalence} and
Theorem~\ref{theorem_main_result} as stated in the next theorem.
\begin{Theo}
\label{cor_main_result} The capacity region of the Gaussian MIMO
wiretap channel with public messages $\mathcal{C}_p(\bbs)$ is
given by the union of rate triples $(R_0,R_p,R_s)$ satisfying
\begin{align}
0\leq R_s&\leq
\frac{1}{2}\log\frac{|\bbh_Y\bbk\bbh_Y^\top+\bbsigma_Y|}{|\bbsigma_Y|}-\frac{1}{2}\log\frac{|\bbh_Z\bbk\bbh_Z^\top+\bbsigma_Z|}{|\bbsigma_Z|}
\\
R_0+R_p+R_s&\leq
\frac{1}{2}\log\frac{|\bbh_Y\bbk\bbh_Y^\top+\bbsigma_Y|}{|\bbsigma_Y|}
\nonumber\\
&\quad
+\min\left\{\frac{1}{2}\log\frac{|\bbh_Y\bbs\bbh_Y^\top+\bbsigma_Y|}{|\bbh_Y\bbk\bbh_Y^\top+\bbsigma_Y|},\frac{1}{2}\log\frac{|\bbh_Z\bbs\bbh_Z^\top+\bbsigma_Z|}{|\bbh_Z\bbk\bbh_Z^\top+\bbsigma_Z|}\right\}
\\
R_0&\leq
\min\left\{\frac{1}{2}\log\frac{|\bbh_Y\bbs\bbh_Y^\top+\bbsigma_Y|}{|\bbh_Y\bbk\bbh_Y^\top+\bbsigma_Y|},\frac{1}{2}\log\frac{|\bbh_Z\bbs\bbh_Z^\top+\bbsigma_Z|}{|\bbh_Z\bbk\bbh_Z^\top+\bbsigma_Z|}\right\}
\end{align}
for some positive semi-definite matrix $\bbk$ such that
$\bzero\preceq \bbk\preceq \bbs$.
\end{Theo}

We next define a sub-class of Gaussian MIMO wiretap channels
called the aligned Gaussian MIMO wiretap channel, which can be
obtained from
(\ref{general_gaussian_mimo_1})-(\ref{general_gaussian_mimo_2}) by
setting $\bbh_Y=\bbh_Z=\bbi$,
\begin{align}
\bby&=\bbx+\bbn_Y\label{aligned_channel_1}\\
\bbz&=\bbx+\bbn_Z \label{aligned_channel_2}
\end{align}
In this work, we first prove Theorems~\ref{theorem_main_result}
and~\ref{cor_main_result} for the aligned Gaussian MIMO wiretap
channel. Then, we establish the capacity region for the general
channel model in
(\ref{general_gaussian_mimo_1})-(\ref{general_gaussian_mimo_2}) by
following the analysis in Section~V.B of~\cite{Shamai_MIMO} and
Section~7.1 of~\cite{MIMO_BC_Secrecy} in conjunction with the
capacity result we obtain for the aligned channel.

\subsection{Capacity Region under a Power Constraint}

We note that the covariance constraint on the channel input in
(\ref{covariance_constraint}) is a rather general constraint that
subsumes the average power constraint
\begin{align}
E\left[\bbx^{\top}\bbx\right]={\rm tr}\left(E\left[\bbx
\bbx^{\top}\right]\right)\leq P \label{trace_constraint}
\end{align}
as a special case, see Lemma~1 and Corollary~1
of~\cite{Shamai_MIMO}. Therefore, using
Theorem~\ref{theorem_main_result}, the capacity-equivocation
region arising from the average power constraint in
(\ref{trace_constraint}), $\mathcal{C}(P)$, can be found as
follows.
\begin{Cor}
\label{cor_total_power_constraint} The capacity-equivocation
region of the Gaussian MIMO wiretap channel subject to an average
power constraint $P$, $\mathcal{C}(P)$, is given by the union of
rate triples $(R_0,R_1,R_e)$ satisfying
\begin{align}
0\leq R_e&\leq
\frac{1}{2}\log\frac{|\bbh_Y\bbk_1\bbh_Y^\top+\bbsigma_Y|}{|\bbsigma_Y|}-\frac{1}{2}\log\frac{|\bbh_Z\bbk_1\bbh_Z^\top+\bbsigma_Z|}{|\bbsigma_Z|}\\
R_0+R_1&\leq
\frac{1}{2}\log\frac{|\bbh_Y\bbk_1\bbh_Y^\top+\bbsigma_Y|}{|\bbsigma_Y|}
\nonumber\\
&\quad
+\min\left\{\frac{1}{2}\log\frac{|\bbh_Y(\bbk_1+\bbk_2)\bbh_Y^\top+\bbsigma_Y|}{|\bbh_Y\bbk_1\bbh_Y^\top+\bbsigma_Y|},\frac{1}{2}\log\frac{|\bbh_Z(\bbk_1+\bbk_2)\bbh_Z^\top+\bbsigma_Z|}{|\bbh_Z\bbk_1\bbh_Z^\top+\bbsigma_Z|}\right\}
\\
R_0&\leq
\min\left\{\frac{1}{2}\log\frac{|\bbh_Y(\bbk_1+\bbk_2)\bbh_Y^\top+\bbsigma_Y|}{|\bbh_Y\bbk_1\bbh_Y^\top+\bbsigma_Y|},\frac{1}{2}\log\frac{|\bbh_Z(\bbk_1+\bbk_2)\bbh_Z^\top+\bbsigma_Z|}{|\bbh_Z\bbk_1\bbh_Z^\top+\bbsigma_Z|}\right\}
\end{align}
for some positive semi-definite matrices $\bbk_1,\bbk_2$ such that
${\rm tr}(\bbk_1+\bbk_2)\leq P$.
\end{Cor}

\section{Proof of Theorem~\ref{theorem_main_result} for the Aligned Case}

Instead of proving Theorem~\ref{theorem_main_result}, here we
prove Theorem~\ref{cor_main_result}, which implies
Theorem~\ref{theorem_main_result} due to
Lemma~\ref{lemma_equivalence}. Achievability of the region given
in Theorem~\ref{cor_main_result} can be shown by setting $V=\bbx$
in Theorem~\ref{theorem_csiszar_alternative}, and using jointly
Gaussian $(U,\bbx=U+T)$, where $U,T$ are independent Gaussian
random vectors with covariance matrices $\bbs-\bbk,\bbk$,
respectively. In the rest of this section, we provide the converse
proof. To this end, we note that since $\mathcal{C}_p(\bbs)$ is
convex by definition, it can be characterized by solving the
following optimization problem\footnote{Although characterizing
$\mathcal{C}_p(\bbs)$ by solving the following optimization
problem
\begin{align}
\max_{(R_0,R_p,R_s)\in\mathcal{C}_p(\bbs)} \mu_0R_0+\mu_pR_p+\mu_s
R_s
\end{align}
for all $\mu_0,\mu_p,\mu_s$ seems to be more natural, we find
working with (\ref{generic optimization}) more convenient. Here,
we characterize $\mathcal{C}_p(\bbs)$ by solving (\ref{generic
optimization}) for all $\mu_p,\mu_s,$ for all fixed feasible
$R_0^*$.}
\begin{align}
f(R_0^*)=\max_{(R_0^*,R_p,R_s)\in\mathcal{C}_p(\bbs)}~~\mu_p
R_p+\mu_s R_s \label{generic optimization}
\end{align}
for all $\mu_p\in[0,\infty),\mu_s\in[0,\infty)$, and all possible
common message rates $R_0^*$, which is bounded as follows
\begin{align}
0\leq R_0^* \leq \min\{C_Y(\bbs),C_Z(\bbs)\}
\end{align}
where $C_Y(\bbs),C_Z(\bbs)$ are the single-user capacities for the
legitimate user and the eavesdropper channels, respectively, i.e.,
\begin{align}
C_Y(\bbs)&=\frac{1}{2}\log\frac{|\bbs+\bbsigma_Y|}{|\bbsigma_Y|}\\
C_Z(\bbs)&=\frac{1}{2}\log\frac{|\bbs+\bbsigma_Z|}{|\bbsigma_Z|}
\end{align}
We note that the optimization problem in (\ref{generic
optimization}) can be expressed in the following more explicit
form
\begin{align}
f(R_0^*)=&\max_{\substack{U\rightarrow V\rightarrow \bbx
\rightarrow (\bby,\bbz)\\E\left[\bbx \bbx^\top\right] \preceq
\bbs}}~~\mu_p R_p+\mu_s
R_s \label{generic_optimization_1_1}\\
&\qquad~~ {\rm s.t.}
\left\{
\begin{array}{rcl}
0\leq R_s &\leq &I(V;\bby|U)-I(V;\bbz|U) \\
R_0^*+R_p+R_s &\leq & I(V;\bby|U)+\min\{I(U;\bby),I(U;\bbz)\}\\
R_0^* &\leq & \min\{I(U;\bby),I(U;\bbz)\}
\end{array}\right.
\label{generic_optimization_1_2}
\end{align}
We also consider the Gaussian rate region $\mathcal{R}^G(\bbs)$
which is defined as
\begin{align}
\mathcal{R}^{G}(\bbs)=\left\{ (R_0,R_p,R_s):
\begin{array}{rcl}
0\leq R_s&\leq &R_s(\bbk) \\
R_0+R_p+R_s &\leq &
R_s(\bbk)+R_p(\bbk)+\min\{R_{0Y}(\bbk),R_{0Z}(\bbk)\}\\
R_0 &\leq & \min\{R_{0Y}(\bbk),R_{0Z}(\bbk)\}\\
&&\hspace{-2cm}{\rm for~some~}\bzero\preceq \bbk  \preceq \bbs
\end{array}\right\}
\end{align}
where $R_s(\bbk),R_p(\bbk),R_{0Y}(\bbk),R_{0Z}(\bbk)$ are given as
follows
\begin{align}
R_s(\bbk)&=\frac{1}{2}\log\frac{|\bbk+\bbsigma_Y|}{|\bbsigma_Y|}-\frac{1}{2}\log\frac{|\bbk+\bbsigma_Z|}{|\bbsigma_Z|}\\
R_p(\bbk)&=\frac{1}{2}\log\frac{|\bbk+\bbsigma_Z|}{|\bbsigma_Z|}\\
R_{0Y}(\bbk)&=\frac{1}{2}\log\frac{|\bbs+\bbsigma_Y|}{|\bbk+\bbsigma_Y|}\\
R_{0Z}(\bbk)&=\frac{1}{2}\log\frac{|\bbs+\bbsigma_Z|}{|\bbk+\bbsigma_Z|}
\end{align}
To provide the converse proof, i.e., to prove the optimality of
jointly Gaussian $(U,V=\bbx)$ for the optimization problem in
(\ref{generic_optimization_1_1})-(\ref{generic_optimization_1_2}),
we will show that
\begin{align}
f(R_0^*)=g(R_0^*),\quad 0\leq R_0^*\leq
\min\{C_Y(\bbs),C_Z(\bbs)\} \label{ultimate_goal}
\end{align}
where $g(R_0^*)$ is defined as
\begin{align}
g(R_0^*)=\max_{(R_0^*,R_p,R_s)\in\mathcal{R}^{G}(\bbs)}~~ \mu_p
R_p+\mu_s R_s \label{optimization_Gaussian}
\end{align}
We show (\ref{ultimate_goal}) in two parts:
\begin{itemize}
\item $\mu_s\leq \mu_p$ \item $\mu_p<\mu_s$
\end{itemize}

\subsection{$\mu_s\leq \mu_p$}
In this case, $f(R_0^*)$ can be written as
\begin{align}
f(R_0^*)=&\max_{\substack{U\rightarrow V\rightarrow \bbx
\rightarrow (\bby,\bbz)\\E\left[\bbx \bbx^\top\right] \preceq
\bbs}}~~\mu_p
(R_p+R_s) \label{generic_optimization_2_1}\\
&\qquad~~ {\rm s.t.} \left\{
\begin{array}{rcl}
R_0^*+R_p+R_s &\leq & I(\bbx;\bby|U)+\min\{I(U;\bby),I(U;\bbz)\}\\
R_0^* &\leq & \min\{I(U;\bby),I(U;\bbz)\}
\end{array}\right.
\label{generic_optimization_2_2}
\end{align}
where we use the fact that $\mu_s\leq \mu_p$, and the secret
message rate $R_s$ can be given up in favor of the private message
rate $R_p$. This optimization problem gives us the capacity region
of the two-user Gaussian MIMO broadcast channel with degraded
message sets, where a common message is sent to both users, and a
private message, on which there is no secrecy constraint, is sent
to one of the two users~\cite{Marton_Deg_Message_Set}. The
optimization problem for this case given in
(\ref{generic_optimization_2_1})-(\ref{generic_optimization_2_2})
is solved in~\cite{Hannan_Common} by showing the optimality of
jointly Gaussian $(U,\bbx)$, i.e., $f(R_0^*)=g(R_0^*)$. This
completes the converse proof for the case $\mu_s\leq\mu_p$.

\subsection{$\mu_p< \mu_s$}
In this case, we first study the optimization problem in
(\ref{optimization_Gaussian}). We rewrite $g(R_0^*)$ as follows
\begin{align}
g(R_0^*)=&\max_{\substack{\bzero\preceq \bbk\preceq
\bbs\\R_p}}~~\mu_p
R_p+\mu_s R_s(\bbk) \label{optimization_Gaussian_again_1}\\
&~~~ {\rm s.t.}~~~\left\{
\begin{array}{rcl}
R_0^*+R_p &\leq & R_p(\bbk)+\min\{R_{0Y}(\bbk),R_{0Z}(\bbk)\}\\
R_0^* &\leq & \min\{R_{0Y}(\bbk),R_{0Z}(\bbk)\}
\end{array}\right.
\label{optimization_Gaussian_again_2}
\end{align}
Let $(\bbk^*,R_p^*)$ be the maximizer for this optimization
problem. The necessary KKT conditions that $(\bbk^*,R_p^*)$ needs
to satisfy are given in the following lemma.
\begin{Lem}
\label{lemma_KKT_conditions} $\bbk^*$ needs to satisfy
\begin{align}
(\mu_s-\mu_p\lambda-\beta_Y)(\bbk^*+\bbsigma_Y)^{-1}+\bbm
&=(\mu_s-\mu_p \lambda+\beta_Z)(\bbk^*+\bbsigma_Z)^{-1}+\bbm_S
\label{KKT_1}
\end{align}
for some positive semi-definite matrices $\bbm, \bbm_S$ such that
\begin{align}
\bbk^*\bbm&=\bbm\bbk^*=\bzero \label{KKT_2}\\
(\bbs-\bbk^*)\bbm_S &=\bbm_S(\bbs-\bbk^*)=\bzero\label{KKT_3}
\end{align}
and for some $\lambda=1-\bar{\lambda}$ such that it satisfies
$0\leq \lambda\leq 1$ and
\begin{align}
\lambda\left\{
\begin{array}{rcl}
&=&0\qquad {\rm if}\quad R_{0Y}(\bbk^*)>R_{0Z}(\bbk^*)\\
&=&1 \qquad {\rm if} \quad R_{0Y}(\bbk^*)<R_{0Z}(\bbk^*)\\
&\neq & 0,1 \quad  {\rm if} \quad R_{0Y}(\bbk^*)=R_{0Z}(\bbk^*)
\end{array}
\right. \label{KKT_4}
\end{align}
and $(\beta_Y,\beta_Z)$ are given as follows
\begin{align}
(\beta_Y,\beta_Z)=\left\{
\begin{array}{rcl}
&(0,0)&\qquad {\rm if}\quad R_0^*<\min\{R_{0Y}(\bbk^*),R_{0Z}(\bbk^*)\}\\
& (0,>0)& \qquad {\rm if} \quad R_0^*=R_{0Z}(\bbk^*)<R_{0Y}(\bbk^*)\\
& (>0,0)& \qquad {\rm if} \quad R_0^*=R_{0Y}(\bbk^*)<R_{0Z}(\bbk^*)\\
&  (>0,>0)& \qquad  {\rm if} \quad R_0^*=
R_{0Y}(\bbk^*)=R_{0Z}(\bbk^*)
\end{array}
\right. \label{KKT_5}
\end{align}
$R_p^*$ needs to satify
\begin{align}
R_p^*=R_p(\bbk^*)+\min\{R_{0Y}(\bbk^*),R_{0Z}(\bbk^*)\}-R_0^*
\label{KKT_6}
\end{align}
\end{Lem}
The proof of Lemma~\ref{lemma_KKT_conditions} is given in
Appendix~\ref{proof_of_lemma_KKT_conditions}. We treat three cases
separately:
\begin{itemize}
\item $R_0^*<\min\{R_{0Y}(\bbk^*),R_{0Z}(\bbk^*)\}$ \item
$R_0^*=R_{0Y}(\bbk^*)\leq R_{0Z}(\bbk^*)$ \item
$R_0^*=R_{0Z}(\bbk^*)<R_{0Y}(\bbk^*)$
\end{itemize}

\subsubsection{$R_0^*<\min\{R_{0Y}(\bbk^*),R_{0Z}(\bbk^*)\}$ }
\label{R0_very_small} In this case, we have $\beta_Y=\beta_Z=0$,
see (\ref{KKT_5}). Thus, the KKT condition in (\ref{KKT_1})
reduces to
\begin{align}
(\mu_s-\mu_p\lambda)(\bbk^*+\bbsigma_Y)^{-1}+\bbm &=(\mu_s-\mu_p
\lambda)(\bbk^*+\bbsigma_Z)^{-1}+\bbm_S \label{KKT_1_again}
\end{align}
We first note that $\bbk^*$ satisfying (\ref{KKT_1_again})
achieves the secrecy capacity of this Gaussian MIMO wiretap
channel~\cite{Tie_Liu_MIMO_WT}, i.e.,
\begin{align}
R_s^*&=R_s(\bbk^*)\label{i_need_u}\\
&=C_S(\bbs)\\
&=\max_{\bzero\preceq \bbk\preceq
\bbs}\frac{1}{2}\log\frac{|\bbk+\bbsigma_Y|}{|\bbsigma_Y|}-\frac{1}{2}\log\frac{|\bbk+\bbsigma_Z|}{|\bbsigma_Z|}
\label{achieves_Cs}
\end{align}
Next, we define a new covariance matrix $\tilde{\bbsigma}_Z$ as
follows
\begin{align}
(\mu_s-\mu_p\lambda)(\bbk^*+\tilde{\bbsigma}_Z)^{-1}=(\mu_s-\mu_p\lambda)(\bbk^*+\bbsigma_Z)^{-1}+\bbm_S
\label{dummy_enhancement}
\end{align}
This new covariance matrix $\tilde{\bbsigma}_Z$ has some useful
properties which are listed in the following lemma.
\begin{Lem}
\label{lemma_dummy_enhancement} We have the following facts.
\begin{itemize}
\item $\tilde{\bbsigma}_Z \preceq \bbsigma_Z$ \item
$\tilde{\bbsigma}_Z \preceq \bbsigma_Y$ \item
$(\bbk^*+\tilde{\bbsigma}_Z)^{-1}(\bbs+\tilde{\bbsigma}_Z)=(\bbk^*+\bbsigma_Z)^{-1}(\bbs+\bbsigma_Z)$
\end{itemize}
\end{Lem}
The proof of Lemma~\ref{lemma_dummy_enhancement} is given in
Appendix~\ref{proof_of_dummy_enhancement}. Thus, we have
\begin{align}
R_{0Z}(\bbk^*)&=\frac{1}{2}\log\frac{|\bbs+\bbsigma_Z|}{|\bbk^*+\bbsigma_Z|} \\
&=\frac{1}{2}\log\frac{|\bbs+\tilde{\bbsigma}_Z|}{|\bbk^*+\tilde{\bbsigma}_Z|} \label{dummy_enhancement_implies} \\
&\geq \frac{1}{2}\log\frac{|\bbs+\bbsigma_Y|}{|\bbk^*+\bbsigma_Y|}
\label{monotonicity_implies} \\
&=R_{0Y}(\bbk^*)
\end{align}
where (\ref{dummy_enhancement_implies}) comes from the third part
of Lemma~\ref{lemma_dummy_enhancement},
(\ref{monotonicity_implies}) is due to the fact that
\begin{align}
\frac{|\bba+\bbb+\bbdelta|}{|\bbb+\bbdelta|}\leq
\frac{|\bba+\bbb|}{|\bbb|}
\end{align}
for $\bba\succeq \bzero$, $\bbdelta\succeq \bzero$,
$\bbb\succ\bzero$ by noting the second part of
Lemma~\ref{lemma_dummy_enhancement}. Therefore, we have
\begin{align}
R_{0Z}(\bbk^*)\geq
R_{0Y}(\bbk^*)\label{the_most_dummy_enhancement_implies}
\end{align}
where $\bbk^*$ satisfies (\ref{KKT_1_again}). Using
(\ref{the_most_dummy_enhancement_implies}) in (\ref{KKT_6}), we
find $R_p^*$ as follows
\begin{align}
R_p^*=R_p(\bbk^*)+R_{0Y}(\bbk^*)-R_0^*
\end{align}
We also note the following
\begin{align}
R_0^*+R_p^*+R_s^*&=R_{0Y}(\bbk^*)+R_p(\bbk^*)+R_{s}(\bbk^*)\label{i_need_u_1}\\
&=\frac{1}{2} \log\frac{|\bbs+\bbsigma_Y|}{|\bbsigma_Y|}\\
&=C_Y(\bbs)\label{achieves_C1}
\end{align}
Now, we show that
\begin{align}
g(R_0^*)=f(R_0^*) \label{to_show}
\end{align}
To this end, we assume that
\begin{align}
g(R_0^*)<f(R_0^*)\label{assumption}
\end{align}
which implies that there exists a rate triple
$(R_0^*,R_p^o,R_s^o)\in\mathcal{C}_p(\bbs)$ such that
\begin{align}
\mu_p R_p^*+\mu_s R_s^*<\mu_p R_p^o+\mu_s R_s^o
\label{assumption_1}
\end{align}
To prove (\ref{to_show}), i.e., that (\ref{assumption}) is not
possible, we note the following bounds
\begin{align}
R_s^o&\leq C_S(\bbs)=R_s^* \label{dummy_bound_1}\\
R_p^o+R_s^o&\leq C_Y(\bbs)-R_0^*=R_p^*+R_s^* \label{dummy_bound_2}
\end{align}
where (\ref{dummy_bound_1}) comes from (\ref{achieves_Cs}) and the
fact that the rate of the confidential message, i.e., $R_s$,
cannot exceed the secrecy capacity, and (\ref{dummy_bound_2}) is
due to (\ref{achieves_C1}) and the fact that the sum rate
$R_0+R_p+R_s$ cannot exceed the legitimate user's single-user
capacity. Thus, in view of $\mu_s>\mu_p$,
(\ref{dummy_bound_1})-(\ref{dummy_bound_2}) imply
\begin{align}
\mu_p R_p^o+\mu_s R_s^o&\leq \mu_p R_p^*+\mu_s R_s^*
\end{align}
which contradicts with (\ref{assumption_1}); proving
(\ref{to_show}). This completes the converse proof for this case.

Before starting the proofs of the other two cases, we now recap
our proof for the case $R_0^*
<\min\{R_{0Y}(\bbk^*),R_{0Z}(\bbk^*)\}$. We note that we did not
show the optimality of Gaussian signalling directly, instead, we
prove it indirectly by showing the following
\begin{align}
g(R_0^*)=f(R_0^*) \label{recap_to_show}
\end{align}
First, we show that for the given common message rate $R_0^*$, we
can achieve the secrecy capacity, i.e., $R_s^*=C_S(\bbs)$, see
(\ref{i_need_u})-(\ref{achieves_Cs}). In other words, we show that
$(R_0^*,0,R_s^*)$ is on the boundary of the capacity region
$\mathcal{C}_p(\bbs)$. Secondly, we show that for the given common
message rate $R_0^*$, $(R_p^*,R_s^*)$ achieve the sum capacity of
the public and confidential messages, i.e., $R_s^*+R_p^*$ is sum
rate optimal for the given common message rate $R_0^*$, see
(\ref{i_need_u_1})-(\ref{achieves_C1}) and (\ref{dummy_bound_2}).
These two findings lead to the inequalities in
(\ref{dummy_bound_1})-(\ref{dummy_bound_2}). Finally, we use a
time-sharing argument for these two inequalities in
(\ref{dummy_bound_1})-(\ref{dummy_bound_2}) to obtain
(\ref{recap_to_show}), which completes the proof.

\subsubsection{$R_0^*=R_{0Y}(\bbk^*)\leq R_{0Z}(\bbk^*)$ }

\label{sec:hard}

We first rewrite the KKT condition in (\ref{KKT_1}) as follows
\begin{align}
(\mu_s-\mu_p\lambda-\mu_0\beta)(\bbk^*+\bbsigma_Y)^{-1}+\bbm=(\mu_s-\mu_p\lambda+\mu_0\bar{\beta})(\bbk^*+\bbsigma_Z)^{-1}+\bbm_S
\label{KKT_1_again_1}
\end{align}
by defining $\mu_0=\beta_Y+\beta_Z$, $\mu_0\beta=\beta_Y$, and
$\mu_0\bar{\beta}=\beta_Z$. We note that if
$R_{0Y}(\bbk^*)<R_{0Z}(\bbk^*)$, we have $\beta=\lambda=1$, if
$R_{0Y}(\bbk^*)=R_{0Z}(\bbk^*)$, we have $0<\lambda<1,0<\beta<1$.
The proof of these two cases are very similar, and we consider
only the case $0<\lambda<1,0<\beta<1$, i.e., we assume
$R_{0Y}(\bbk^*)=R_{0Z}(\bbk^*)$. The other case can be proved
similarly.

Similar to Section~\ref{R0_very_small}, here also, we prove the
desired identity
\begin{align}
g(R_0^*)=f(R_0^*) \label{desired_identity}
\end{align}
by contradiction. We first assume that
\begin{align}
g(R_0^*)<f(R_0^*)\label{assumption_again}
\end{align}
which implies that there exists a rate triple
$(R_0^*,R_p^o,R_s^o)\in\mathcal{C}_p(\bbs)$ such that
\begin{align}
\mu_p R_p^*+\mu_s R_s^*<\mu_p R_p^o+\mu_s R_s^o
\label{assumption_again_1}
\end{align}
where we define $R_s^*=R_s(\bbk^*)$. Since the sum rate
$R_0+R_p+R_s$ needs to be smaller than the legitimate user's
single user capacity, we have
\begin{align}
R_0^*+R_p^o+R_s^o \leq C_Y(\bbs)\label{obvious_bound_again}
\end{align}
On the other hand, we have the following
\begin{align}
R_0^*+R_p^*+R_s^*&=\min\{R_{0Y}(\bbk^*),R_{0Z}(\bbk^*)\}+R_p(\bbk^*)+R_s(\bbk^*)\label{KKT_6_implies}\\
&=R_{0Y}(\bbk^*)+R_p(\bbk^*)+R_s(\bbk^*) \label{we_assume}\\
&=C_Y(\bbs) \label{obvious_bound}
\end{align}
where (\ref{KKT_6_implies}) comes from (\ref{KKT_6}), and
(\ref{we_assume}) is due to our assumption that
$R_0^*=R_{0Y}(\bbk^*)= R_{0Z}(\bbk^*)$. Equations
(\ref{obvious_bound_again}) and (\ref{obvious_bound}) imply that
\begin{align}
R_p^o+R_s^o\leq R_p^*+R_s^* \label{towards_contradiction}
\end{align}

In the rest of this section, we prove that we have $R_s^o\leq
R_s^*$ for the given common message rate $R_0^*$, which, in
conjunction with (\ref{towards_contradiction}), will yield a
contradiction with (\ref{assumption_again_1}); proving
(\ref{desired_identity}). To this end, we first define a new
covariance matrix $\tilde{\bbsigma}_Y$ as follows
\begin{align}
(\mu_s-\mu_p\lambda)(\bbk^*+\tilde{\bbsigma}_Y)^{-1}=(\mu_s-\mu_p\lambda)(\bbk^*+\bbsigma_Y)^{-1}+\bbm
\label{one_more_time}
\end{align}
This new covariance matrix $\tilde{\bbsigma}_Y$ has some useful
properties which are listed in the following lemma.
\begin{Lem}
\label{lemma_one_more_time} We have the following facts.
\begin{itemize}
\item $\tilde{\bbsigma}_Y \preceq \bbsigma_Y$ \item
$\tilde{\bbsigma}_Y \preceq \bbsigma_Z$ \item
$(\bbk^*+\tilde{\bbsigma}_Y)^{-1}\tilde{\bbsigma}_Y=(\bbk^*+\bbsigma_Y)^{-1}\bbsigma_Y$
\end{itemize}
\end{Lem}
The proof of this lemma is given in
Appendix~\ref{proof_of_lemma_one_more_time}. Using this new
covariance matrix, we define a random vector $\tilde{\bby}$ as
\begin{align}
\tilde{\bby}=\bbx+\tilde{\bbn}_Y
\end{align}
where $\tilde{\bbn}_Y$ is a Gaussian random vector with covariance
matrix $\tilde{\bbsigma}_Y$. Due to the first and second
statements of Lemma~\ref{lemma_one_more_time}, we have the
following Markov chains
\begin{align}
&U\rightarrow V\rightarrow \bbx \rightarrow
\tilde{\bby}\rightarrow
\bby \label{really_dummy_MC_1}\\
&U\rightarrow V\rightarrow \bbx \rightarrow
\tilde{\bby}\rightarrow \bbz \label{really_dummy_MC_2}
\end{align}
We next study the following optimization problem
\begin{align}
\lefteqn{\max_{(R_0,R_p,R_s)\in\mathcal{C}_p(\bbs)}~~\mu_0R_0+(\mu_s-\mu_p\lambda)R_s}\nonumber\\
&=\max_{\substack{U\rightarrow V\rightarrow \bbx\rightarrow
(\bby,\bbz)\\E\left[\bbx\bbx^\top\right]\preceq
\bbs}}~~\mu_0\min\{I(U;\bby),I(U;\bbz)\}+(\mu_s-\mu_p\lambda)\left[I(V;\bby|U)-I(V;\bbz|U)\right]
\label{optimization_for_contradiction}
\end{align}
Since we assume $(R_0^*,R_p^o,R_s^o)\in\mathcal{C}_p(\bbs)$, we
have the following lower bound for
(\ref{optimization_for_contradiction})
\begin{align}
\mu_0R_0^*+(\mu_s-\mu_p\lambda) R_s^o\leq
\max_{(R_0,R_p,R_s)\in\mathcal{C}_p(\bbs)}~~\mu_0R_0+(\mu_s-\mu_p\lambda)R_s
\label{optimization_for_contradiction_lower_bound}
\end{align}
Now we solve the optimization problem in
(\ref{optimization_for_contradiction}) as follows
\begin{align}
\lefteqn{\max_{(R_0,R_p,R_s)\in\mathcal{C}_p(\bbs)}~~\mu_0R_0+(\mu_s-\mu_p\lambda)R_s}\nonumber\\
&=\max_{\substack{U\rightarrow V\rightarrow \bbx\rightarrow
(\bby,\bbz)\\E\left[\bbx\bbx^\top\right]\preceq
\bbs}}~~\mu_0\min\{I(U;\bby),I(U;\bbz)\}+(\mu_s-\mu_p\lambda)\left[I(V;\bby|U)-I(V;\bbz|U)\right]\\
&\leq \max_{\substack{U\rightarrow V\rightarrow \bbx\rightarrow
(\bby,\bbz)\\E\left[\bbx\bbx^\top\right]\preceq
\bbs}}~~\mu_0\bar{\beta}I(U;\bbz)+\mu_0\beta
I(U;\bby)+(\mu_s-\mu_p\lambda)\left[I(V;\bby|U)-I(V;\bbz|U)\right]\label{min_is_not_min} \\
&\leq \max_{\substack{U\rightarrow V\rightarrow \bbx\rightarrow
(\bby,\bbz)\\E\left[\bbx\bbx^\top\right]\preceq
\bbs}}~~\mu_0\bar{\beta}I(U;\bbz)+\mu_0\beta
I(U;\bby)+(\mu_s-\mu_p\lambda)\left[I(V;\tilde{\bby}|U)-I(V;\bbz|U)\right]\label{really_dummy_MC_1_implies} \\
&\leq \max_{\substack{U\rightarrow \bbx\rightarrow
(\bby,\bbz)\\E\left[\bbx\bbx^\top\right]\preceq
\bbs}}~~\mu_0\bar{\beta}I(U;\bbz)+\mu_0\beta
I(U;\bby)+(\mu_s-\mu_p\lambda)\left[I(\bbx;\tilde{\bby}|U)-I(\bbx;\bbz|U)\right]\label{really_dummy_MC_2_implies} \\
& \leq
\frac{\mu_0\bar{\beta}}{2}\log\frac{|\bbs+\bbsigma_Z|}{|\bbk^*+\bbsigma_Z|}
+\frac{\mu_0\beta}{2}\log\frac{|\bbs+\bbsigma_Y|}{|\bbk^*+\bbsigma_Y|}+\frac{\mu_s-\mu_p\lambda}{2}\left[\log\frac{|\bbk^*+\tilde{\bbsigma}_Y|}{|\tilde{\bbsigma}_Y|}-\log\frac{|\bbk^*+\bbsigma_Z|}{|\bbsigma_Z|}\right]
\label{liu_extremal_implies}\\
&=\mu_0\bar{\beta}R_{0Z}(\bbk^*)
+\mu_0\beta R_{0Y}(\bbk^*)+\frac{\mu_s-\mu_p\lambda}{2}\left[\log\frac{|\bbk^*+\tilde{\bbsigma}_Y|}{|\tilde{\bbsigma}_Y|}-\log\frac{|\bbk^*+\bbsigma_Z|}{|\bbsigma_Z|}\right]\\
&=\mu_0\bar{\beta}R_{0Z}(\bbk^*)
+\mu_0\beta R_{0Y}(\bbk^*)+\frac{\mu_s-\mu_p\lambda}{2}\left[\log\frac{|\bbk^*+\bbsigma_Y|}{|\bbsigma_Y|}-\log\frac{|\bbk^*+\bbsigma_Z|}{|\bbsigma_Z|}\right] \label{lemma_one_more_time_implies}\\
&=\mu_0\bar{\beta}R_{0Z}(\bbk^*)
+\mu_0\beta R_{0Y}(\bbk^*)+(\mu_s-\mu_p\lambda)R_s(\bbk^*)\\
&=\mu_0 R_0^*+(\mu_s-\mu_p\lambda)R_s^*
\label{close_to_contradiction}
\end{align}
where (\ref{min_is_not_min}) comes from the fact that
$0<\beta=1-\bar{\beta}<1$,
(\ref{really_dummy_MC_1_implies})-(\ref{really_dummy_MC_2_implies})
are due to the Markov chains in
(\ref{really_dummy_MC_1})-(\ref{really_dummy_MC_2}), respectively,
(\ref{liu_extremal_implies}) can be obtained by using the analysis
in~\cite[eqns (30)-(32)]{Liu_Common_Confidential},
(\ref{lemma_one_more_time_implies}) comes from the third part of
Lemma~\ref{lemma_one_more_time}, and
(\ref{close_to_contradiction}) is due to our assumption that
$R_0^*=R_{0Y}(\bbk^*)=R_{0Z}(\bbk^*)$. Thus,
(\ref{close_to_contradiction}) implies
\begin{align}
\max_{(R_0,R_p,R_s)\in\mathcal{C}_p(\bbs)}~~\mu_0R_0+(\mu_s-\mu_p\lambda)R_s\leq
\mu_0 R_0^*+(\mu_s-\mu_p\lambda) R_s^*
\label{close_to_contradiction_dummy}
\end{align}
Comparing (\ref{optimization_for_contradiction_lower_bound}) and
(\ref{close_to_contradiction_dummy}) yields
\begin{align}
R_s^o\leq R_s^* \label{close_to_contradiction_1}
\end{align}
Using (\ref{towards_contradiction}) and
(\ref{close_to_contradiction_1}) and noting $\mu_s>\mu_p$, we can
get
\begin{align}
\mu_pR_p^o+\mu_sR_s^o\leq \mu_pR_p^*+\mu_sR_s^*
\end{align}
which contradicts with (\ref{assumption_again_1}); proving
(\ref{desired_identity}). This completes the converse proof for
this case.

Before providing the proof for the last case, we recap our proof
for the case $R_0^*=R_{0Y}(\bbk^*)\leq R_{0Z}(\bbk^*)$. Similar to
Section~\ref{R0_very_small}, here also, we prove the optimality of
Gaussian signalling indirectly, i.e., we show the desired identity
\begin{align}
g(R_0^*)=f(R_0^*)\label{recap_to_show_1}
\end{align}
indirectly. First, we show that for the given common message rate
$R_0^*$, $R_s^*+R_p^*$ is sum rate optimal, i.e., $(R_p^*,R_s^*)$
achieve the sum capacity of the public and confidential messages,
by obtaining (\ref{towards_contradiction}). Secondly, we show that
$(R_0^*,0,R_s^*)$ is also on the boundary of the capacity region
$\mathcal{C}_p(\bbs)$ by obtaining
(\ref{close_to_contradiction_dummy}). These two findings give us
the inequalities in (\ref{towards_contradiction}) and
(\ref{close_to_contradiction_1}). Finally, we use a time-sharing
argument for these two inequalities in
(\ref{towards_contradiction}) and (\ref{close_to_contradiction_1})
to establish (\ref{recap_to_show_1}), which completes the proof.

\subsubsection{$R_0^*=R_{0Z}(\bbk^*)<R_{0Y}(\bbk^*)$}
In this case, we have $\lambda=\beta_Y=0$, see
(\ref{KKT_4})-(\ref{KKT_5}). Hence, the KKT condition in
(\ref{KKT_1}) reduces to
\begin{align}
\mu_s (\bbk^*+\bbsigma_Y)^{-1}+\bbm&=
(\mu_s+\beta_Z)(\bbk^*+\bbsigma_Z)^{-1}+\bbm_S
\label{KKT_1_again_2}
\end{align}
We again prove the desired identity
\begin{align}
g(R_0^*)=f(R_0^*) \label{desired_identity_again}
\end{align}
by contradiction. We first assume that
\begin{align}
g(R_0^*)<f(R_0^*)\label{assumption_again_2}
\end{align}
which implies that there exists a rate triple
$(R_0^*,R_p^o,R_s^o)\in\mathcal{C}_p(\bbs)$ such that
\begin{align}
\mu_p R_p^*+\mu_s R_s^*<\mu_p R_p^o+\mu_s R_s^o
\label{assumption_again_3}
\end{align}
In the rest of the section, we show that
\begin{align}
\mu_p R_p^*+\mu_s R_s^*\geq \mu_p R_p^o+\mu_s R_s^o
\end{align}
to reach a contradiction, and hence, prove
(\ref{desired_identity_again}). To this end, we define a new
covariance matrix $\tilde{\bbsigma}_Y$ as follows
\begin{align}
\mu_s (\bbk^*+\tilde{\bbsigma}_Y)^{-1}=\mu_s
(\bbk^*+\bbsigma_Y)^{-1}+\bbm \label{enhancement_final}
\end{align}
This new covariance matrix $\tilde{\bbsigma}_Y$ has some useful
properties listed in the following lemma.
\begin{Lem}
\label{lemma_enhancement} We have the following facts.
\begin{itemize}
\item $\tilde{\bbsigma}_Y\preceq \bbsigma_Y$ \item
$\tilde{\bbsigma}_Y\preceq \bbsigma_Z$ \item
$(\bbk^*+\tilde{\bbsigma}_Y)^{-1}\tilde{\bbsigma}_Y=(\bbk^*+
\bbsigma_Y)^{-1}\bbsigma_Y$
\end{itemize}
\end{Lem}
The proof of this lemma is very similar to the proof
Lemma~\ref{lemma_one_more_time}, and hence is omitted. Using this
new covariance matrix $\tilde{\bbsigma}_Y$, we define a random
vector $\tilde{\bby}$ as
\begin{align}
\tilde{\bby}=\bbx+\tilde{\bbn}_Y
\end{align}
where $\tilde{\bbn}_Y$ is a Gaussian random vector with covariance
matrix $\tilde{\bbsigma}_Y$. Due to the first and second
statements of Lemma~\ref{lemma_enhancement}, we have the following
Markov chains
\begin{align}
&U\rightarrow V\rightarrow \bbx \rightarrow
\tilde{\bby}\rightarrow
\bby \label{really_dummy_MC_again_1}\\
&U\rightarrow V\rightarrow \bbx \rightarrow
\tilde{\bby}\rightarrow \bbz \label{really_dummy_MC_again_2}
\end{align}
Next, we study the following optimization problem
\begin{align}
\max_{(R_0,R_p,R_s)\in\mathcal{C}_p(\bbs)}~~(\mu_p+\beta_Z)R_0+\mu_pR_p+\mu_s
R_s \label{optimization_for_contradiction_again}
\end{align}
We note that since $(R_0^*,R_p^o,R_s^o)\in\mathcal{C}_p(\bbs)$, we
have the following lower bound for the optimization problem in
(\ref{optimization_for_contradiction_again})
\begin{align}
(\mu_p+\beta_Z) R_0^*+\mu_pR_p^o+\mu_sR_s^o\leq
\max_{(R_0,R_p,R_s)\in\mathcal{C}_p(\bbs)}~~(\mu_p+\beta_Z) R_0
+\mu_pR_p+\mu_sR_s\label{optimization_for_contradiction_again_1_lower_bound}
\end{align}
We next obtain the maximum for
(\ref{optimization_for_contradiction_again}). To this end, we
introduce the following lemma which provides an explicit form for
this optimization problem.
\begin{Lem}For $\mu_s>\mu_p$, we have
\label{lemma_explicit_form}
\begin{align}
\lefteqn{\hspace{-1.75cm}\max_{(R_0,R_p,R_s)\in\mathcal{C}_p(\bbs)}~~(\mu_p+\beta_Z)
R_0+\mu_pR_p+\mu_sR_s}\nonumber\\
&=\max_{\substack{U\rightarrow V\rightarrow \bbx \rightarrow
(\bby,\bbz)\\E\left[\bbx\bbx^\top\right]\preceq
\bbs}}~~(\mu_p+\beta_Z)\min\{I(U;\bby),I(U;\bbz)\}+\mu_p
I(V;\bbz|U)\nonumber\\
&\qquad\qquad \qquad \qquad +\mu_s
\left[I(V;\bby|U)-I(V;\bbz|U)\right]
\label{optimization_for_contradiction_again_1}
\end{align}
\end{Lem}
The proof of this lemma is given in
Appendix~\ref{proof_of_lemma_explicit_form}.

Next we introduce the following extremal inequality
from~\cite{Liu_Compound}, which will be used subsequently in the
solution of (\ref{optimization_for_contradiction_again_1}).
\begin{Lem}{\bf (\!\cite[Corollary~4]{Liu_Compound})}
\label{lemma_extremal_liu} Let $(U,\bbx)$ be an arbitrarily
correlated random vector, where $\bbx$ has a covariance constraint
$E\left[\bbx\bbx^{\top}\right] \preceq \bbs$ and $\bbs\succ
\bzero$. Let $\bbn_1,\bbn_2$ be Gaussian random vectors with
covariance matrices $\bbsigma_1,\bbsigma_2$, respectively. They
are independent of $(U,\bbx)$. Furthermore,
$\bbsigma_1,\bbsigma_2$ satisfy $\bbsigma_1\preceq \bbsigma_2$.
Assume that there exists a covariance matrix $\bbk^*$ such that
$\bbk^*\preceq \bbs$ and
\begin{align}
\nu(\bbk^*+\bbsigma_1)^{-1}=\gamma (\bbk^*+\bbsigma_2)^{-1}+\bbm_S
\end{align}
where $\nu\geq 0, \gamma\geq 0$ and $\bbm_S$ is positive
semi-definite matrix such that $(\bbs-\bbk^*)\bbm_S=\bzero$. Then,
for any $(U,\bbx)$, we have
\begin{align}
\nu h(\bbx+\bbn_1|U)-\gamma h(\bbx+\bbn_2|U) \leq \frac{\nu}{2}
\log |(2\pi e)(\bbk^*+\bbsigma_1)|-\frac{\gamma}{2} \log |(2\pi
e)(\bbk^*+\bbsigma_2)|
\end{align}
\end{Lem}

Now we use Lemma~\ref{lemma_extremal_liu}. To this end, we note
that using (\ref{enhancement_final}) in (\ref{KKT_1_again_2}), we
get
\begin{align}
\mu_s(\bbk^*+\tilde{\bbsigma}_Y)^{-1}=(\mu_s+\beta_Z)(\bbk^*+\bbsigma_Z)^{-1}+\bbm_S
\label{KKT_1_after_enhancement_1}
\end{align}
In view of (\ref{KKT_1_after_enhancement_1}) and the fact that
$\tilde{\bbsigma}_Y\preceq \bbsigma_Z$,
Lemma~\ref{lemma_extremal_liu} implies
\begin{align}
\mu_s h(\tilde{\bby}|U)-(\mu_s+\beta_Z)h(\bbz|U)\leq
\frac{\mu_s}{2} \log |(2\pi
e)(\bbk^*+\tilde{\bbsigma}_Y)|-\frac{\mu_s+\beta_Z}{2}\log|(2\pi
e)(\bbk^*+\bbsigma_Z)| \label{exremal_inequality_implies_1}
\end{align}
We now consider the maximization in
(\ref{optimization_for_contradiction_again_1}) as follows
\begin{align}
\lefteqn{\max_{(R_0,R_p,R_s)\in\mathcal{C}_p(\bbs)}~~(\mu_p+\beta_Z)
R_0+\mu_pR_p+\mu_sR_s}\nonumber\\
&=\max_{\substack{U\rightarrow V\rightarrow \bbx \rightarrow
(\bby,\bbz)\\E\left[\bbx\bbx^\top\right]\preceq
\bbs}}~~(\mu_p+\beta_Z)\min\{I(U;\bby),I(U;\bbz)\}+\mu_p
I(V;\bbz|U)\nonumber\\
&\qquad\qquad \qquad \qquad +\mu_s
\left[I(V;\bby|U)-I(V;\bbz|U)\right]\\
&\leq \max_{\substack{U\rightarrow V\rightarrow \bbx \rightarrow
(\bby,\bbz)\\E\left[\bbx\bbx^\top\right]\preceq
\bbs}}~~(\mu_p+\beta_Z)I(U;\bbz)+\mu_p I(V;\bbz|U)+\mu_s
\left[I(V;\bby|U)-I(V;\bbz|U)\right]
\label{min_is_not_min_again}\\
&\leq \max_{\substack{U\rightarrow V\rightarrow \bbx \rightarrow
(\bby,\bbz)\\E\left[\bbx\bbx^\top\right]\preceq
\bbs}}~~(\mu_p+\beta_Z)I(U;\bbz)+\mu_p I(\bbx;\bbz|U)+\mu_s
\left[I(V;\bby|U)-I(V;\bbz|U)\right]
 \label{some_MC}\\
&\leq \max_{\substack{U\rightarrow V\rightarrow \bbx \rightarrow
(\bby,\bbz)\\E\left[\bbx\bbx^\top\right]\preceq
\bbs}}~~(\mu_p+\beta_Z)I(U;\bbz)+\mu_p I(\bbx;\bbz|U) +\mu_s
\left[I(V;\tilde{\bby}|U)-I(V;\bbz|U)\right]
\label{really_dummy_MC_again_1_implies}\\
&\leq \max_{\substack{U \rightarrow \bbx \rightarrow
(\bby,\bbz)\\E\left[\bbx\bbx^\top\right]\preceq
\bbs}}~~(\mu_p+\beta_Z)I(U;\bbz)+\mu_p I(\bbx;\bbz|U)+\mu_s
\left[I(\bbx;\tilde{\bby}|U)-I(\bbx;\bbz|U)\right]
\label{really_dummy_MC_again_2_implies}
\\
&= \max_{\substack{U \rightarrow \bbx \rightarrow
(\bby,\bbz)\\E\left[\bbx\bbx^\top\right]\preceq
\bbs}}~~(\mu_p+\beta_Z)h(\bbz)+\mu_s
h(\tilde{\bby}|U)-(\mu_s+\beta_Z)h(\bbz|U)\nonumber\\
&\qquad \qquad \qquad \quad -\frac{\mu_s}{2}\log|(2\pi
e)\tilde{\bbsigma}_Y|+\frac{\mu_s-\mu_p}{2}\log|(2\pi
e)\bbsigma_Z|\\
&\leq \frac{\mu_p+\beta_Z}{2}\log |(2\pi
e)(\bbs+\bbsigma_Z)|+\max_{\substack{U \rightarrow \bbx
\rightarrow (\bby,\bbz)\\E\left[\bbx\bbx^\top\right]\preceq
\bbs}}~~\mu_s
h(\tilde{\bby}|U)-(\mu_s+\beta_Z)h(\bbz|U)\nonumber\\
& \quad-\frac{\mu_s}{2}\log|(2\pi
e)\tilde{\bbsigma}_Y|+\frac{\mu_s-\mu_p}{2}\log|(2\pi
e)\bbsigma_Z| \label{max_entropy_thm_implies_1}\\
&\leq \frac{\mu_p+\beta_Z}{2}\log |(2\pi
e)(\bbs+\bbsigma_Z)|+\frac{\mu_s}{2} \log |(2\pi
e)(\bbk^*+\tilde{\bbsigma}_Y)|-\frac{\mu_s+\beta_Z}{2}\log|(2\pi
e)(\bbk^*+\bbsigma_Z)|\nonumber\\
&\quad -\frac{\mu_s}{2}\log|(2\pi
e)\tilde{\bbsigma}_Y|+\frac{\mu_s-\mu_p}{2}\log|(2\pi
e)\bbsigma_Z| \label{exremal_inequality_implies_1_implies}
\\
&=\frac{\mu_p+\beta_Z}{2}\log\frac{|\bbs+\bbsigma_Z|}{|\bbk^*+\bbsigma_Z|}+\frac{\mu_p}{2}\log\frac{|\bbk^*+\bbsigma_Z|}{|\bbsigma_Z|}
+\frac{\mu_s}{2}\left[ \log
\frac{|\bbk^*+\tilde{\bbsigma}_Y|}{|\tilde{\bbsigma}_Y|}- \log
\frac{|\bbk^*+\bbsigma_Z|}{|\bbsigma_Z|}\right]
\\
&=(\mu_p+\beta_Z)R_{0Z}(\bbk^*)+\mu_p R_p(\bbk^*)
+\frac{\mu_s}{2}\left[ \log
\frac{|\bbk^*+\tilde{\bbsigma}_Y|}{|\tilde{\bbsigma}_Y|}- \log
\frac{|\bbk^*+\bbsigma_Z|}{|\bbsigma_Z|}\right]
 \\
&=(\mu_p+\beta_Z)R_{0Z}(\bbk^*)+\mu_p R_p(\bbk^*)
+\frac{\mu_s}{2}\left[ \log
\frac{|\bbk^*+\bbsigma_Y|}{|\bbsigma_Y|}- \log
\frac{|\bbk^*+\bbsigma_Z|}{|\bbsigma_Z|}\right]\label{lemma_enhancement_implies}
\\
&=(\mu_p+\beta_Z)R_{0Z}(\bbk^*)+\mu_p R_p(\bbk^*)
+\mu_sR_s(\bbk^*)\\
&=(\mu_p+\beta_Z)R_0^*+\mu_p R_p^* +\mu_sR_s^*
\label{close_to_contradiction_again}
\end{align}
where (\ref{min_is_not_min_again}) is due to $\min\{a,b\}\leq a$,
(\ref{some_MC}) is due to the Markov chain in
(\ref{really_dummy_MC_again_2}),
(\ref{really_dummy_MC_again_1_implies})-(\ref{really_dummy_MC_again_2_implies})
come from the Markov chains in
(\ref{really_dummy_MC_again_1})-(\ref{really_dummy_MC_again_2}),
respectively, (\ref{max_entropy_thm_implies_1}) is due to the
maximum entropy theorem~\cite{cover_book},
(\ref{exremal_inequality_implies_1_implies}) comes from
(\ref{exremal_inequality_implies_1}), and
(\ref{lemma_enhancement_implies}) is due to the third part of
Lemma~\ref{lemma_enhancement}. Comparing
(\ref{close_to_contradiction_again}) and
(\ref{optimization_for_contradiction_again_1_lower_bound}) yields
\begin{align}
\mu_p R_p^o+\mu_s R_s^o\leq \mu_p R_p^*+\mu_s R_s^*
\end{align}
which contradicts with our assumption in
(\ref{assumption_again_3}); implying
(\ref{desired_identity_again}). This completes the converse proof
for this case.

We note that contrary to Sections~\ref{R0_very_small}
and~\ref{sec:hard}, here we prove the optimality of Gaussian
signalling, i.e.,
\begin{align}
g(R_0^*)=f(R_0^*) \label{recap_to_show_2}
\end{align}
directly. In other words, to show (\ref{recap_to_show_2}), we did
not find any other points on the boundary of the capacity region
$\mathcal{C}_p(\bbs)$ and did not have to use a time-sharing
argument between these points to reach (\ref{recap_to_show_2}).
(This was our strategy in Sections~\ref{R0_very_small}
and~\ref{sec:hard}.) Instead, we define a new optimization problem
given in (\ref{optimization_for_contradiction_again_1}) whose
solution yields (\ref{recap_to_show_2}).

\section{Proof of Theorem~\ref{theorem_main_result} for the General Case}
The achievability of the region given in
Theorem~\ref{theorem_main_result} can be shown by computing the
region in Theorem~\ref{theorem_csiszar} with the following
selection of $(U,V,\bbx)$: $V=\bbx$, $\bbx=\bbu+\bbt$ where
$\bbt,\bbu$ are independent Gaussian random vectors with
covariance matrices $\bbk,\bbs-\bbk$, respectively, $U=\bbu$. In
the rest of this section, we consider the converse proof. We first
note that following the approaches in Section~V.B
of~\cite{Shamai_MIMO} and Section~7.1 of~\cite{MIMO_BC_Secrecy},
it can be shown that a new Gaussian MIMO wiretap channel can be
constructed from any Gaussian MIMO wiretap channel described by
(\ref{general_gaussian_mimo_1})-(\ref{general_gaussian_mimo_2})
such that the new channel has the same capacity-equivocation
region with the original one and in the new channel, both the
legitimate user and the eavesdropper have the same number of
antennas as the transmitter, i.e., $r_Y=r_Z=t$. Thus, without loss
of generality, we assume that $r_Y=r_Z=t$. We next apply
singular-value decomposition to the channel gain matrices
$\bbh_Y,\bbh_Z$ as follows
\begin{align}
\bbh_Y&=\bbu_Y\bblambda_Y\bbv_Y^\top\\
\bbh_Z&=\bbu_Z\bblambda_Z\bbv_Z^\top
\end{align}
where $\bbu_Y,\bbu_Z,\bbv_Y,\bbv_Z$ are $t\times t$ orthogonal
matrices, and $\bblambda_Y,\bblambda_Z$ are diagonal matrices. We
now define a new Gaussian MIMO wiretap channel as follows
\begin{align}
\overline{\bby}&=\overline{\bbh}_Y\bbx+\bbn_Y \label{channel_alpha_1}\\
\overline{\bbz}&=\overline{\bbh}_Z\bbx+\bbn_Z
\label{channel_alpha_2}
\end{align}
where $\overline{\bbh}_Y,\overline{\bbh}_Z$ are defined as
\begin{align}
\overline{\bbh}_Y&=\bbu_Y (\bblambda_Y+\alpha\bbi)\bbv_Y^\top\label{channel_gain_matrix_alpha_1}\\
\overline{\bbh}_Z&=\bbu_Z (\bblambda_Z+\alpha\bbi)\bbv_Z^\top
\label{channel_gain_matrix_alpha_2}
\end{align}
for some $\alpha>0$. We denote the capacity-equivocation region of
the Gaussian MIMO wiretap channel defined in
(\ref{channel_alpha_1})-(\ref{channel_alpha_2}) by
$\mathcal{C}_{\alpha}(\bbs)$. Since
$\overline{\bbh}_Y,\overline{\bbh}_Z$ are invertible, the
capacity-equivocation region of the channel in
(\ref{channel_alpha_1})-(\ref{channel_alpha_2}) is equal to the
capacity-equivocation region of the following aligned channel
\begin{align}
\overline{\overline{\bby}}&=\bbx+\overline{\bbh}_Y^{~-1}\bbn_Y \label{channel_alpha_1_again}\\
\overline{\overline{\bbz}}&=\bbx+\overline{\bbh}_Z^{~-1} \bbn_Z
\label{channel_alpha_2_again}
\end{align}
Thus, using the capacity result for the aligned case, which was
proved in the previous section, we obtain
$\mathcal{C}_{\alpha}(\bbs)$ as the union of rate triples
$(R_0,R_1,R_e)$ satisfying
\begin{align}
0\leq R_e&\leq
\frac{1}{2}\log\frac{\left|\overline{\bbh}_Y\bbk\overline{\bbh}_Y^\top+\bbsigma_Y\right|}{|\bbsigma_Y|}-\frac{1}{2}\log\frac{\left|\overline{\bbh}_Z\bbk\overline{\bbh}_Z^\top+\bbsigma_Z\right|}{|\bbsigma_Z|}
\\
R_0+R_1&\leq
\frac{1}{2}\log\frac{\left|\overline{\bbh}_Y\bbk\overline{\bbh}_Y^\top+\bbsigma_Y\right|}{|\bbsigma_Y|}
\nonumber\\
&\quad
+\min\left\{\frac{1}{2}\log\frac{\left|\overline{\bbh}_Y\bbs\overline{\bbh}_Y^\top+\bbsigma_Y\right|}{\left|\overline{\bbh}_Y\bbk\overline{\bbh}_Y^\top+\bbsigma_Y\right|},\frac{1}{2}\log\frac{\left|\overline{\bbh}_Z\bbs\overline{\bbh}_Z^\top+\bbsigma_Z\right|}{\left|\overline{\bbh}_Z\bbk\overline{\bbh}_Z^\top+\bbsigma_Z\right|}\right\}
\\
R_0&\leq
\min\left\{\frac{1}{2}\log\frac{\left|\overline{\bbh}_Y\bbs\overline{\bbh}_Y^\top+\bbsigma_Y\right|}{\left|\overline{\bbh}_Y\bbk\overline{\bbh}_Y^\top+\bbsigma_Y\right|},\frac{1}{2}\log\frac{\left|\overline{\bbh}_Z\bbs\overline{\bbh}_Z^\top+\bbsigma_Z\right|}{\left|\overline{\bbh}_Z\bbk\overline{\bbh}_Z^\top+\bbsigma_Z\right|}\right\}
\end{align}
for some positive semi-definite matrix $\bbk$ such that
$\bzero\preceq \bbk\preceq \bbs$.

We next obtain an outer bound for the capacity-equivocation region
of the original Gaussian MIMO wiretap channel in
(\ref{general_gaussian_mimo_1})-(\ref{general_gaussian_mimo_2}) in
terms of $\mathcal{C}_\alpha(\bbs)$. To this end, we first note
the following Markov chains
\begin{align}
&\bbx\rightarrow \overline{\bby} \rightarrow \bby
\label{auxiliary_mc_1}\\
&\bbx\rightarrow \overline{\bbz} \rightarrow \bbz
\label{auxiliary_mc_2}
\end{align}
which imply that if the messages $(W_0,W_1)$ with rates
$(R_0,R_1)$ are transmitted with a vanishingly small probability
of error in the original Gaussian MIMO wiretap channel given by
(\ref{general_gaussian_mimo_1})-(\ref{general_gaussian_mimo_2}),
they will be transmitted with a vanishingly small probability of
error in the new Gaussian MIMO wiretap channel given by
(\ref{channel_alpha_1})-(\ref{channel_alpha_2}) as well. However,
as opposed to the rates $R_0,R_1$, we cannot immediately conclude
that if an equivocation rate $R_e$ is achievable in the original
Gaussian MIMO wiretap channel given in
(\ref{general_gaussian_mimo_1})-(\ref{general_gaussian_mimo_2}),
it is also achievable in the new Gaussian MIMO wiretap channel in
(\ref{channel_alpha_1})-(\ref{channel_alpha_2}). The reason for
this is that both the legitimate user's and the eavesdropper's
channel gain matrices are enhanced in the new channel given by
(\ref{channel_alpha_1})-(\ref{channel_alpha_2}), see
(\ref{channel_gain_matrix_alpha_1})-(\ref{channel_gain_matrix_alpha_2})
and/or (\ref{auxiliary_mc_1})-(\ref{auxiliary_mc_2}), and
consequently, it is not clear what the overall effect of these two
enhancements on the equivocation rate will be. However, in the
sequel, we show that if $(R_0,R_1,R_e)\in\mathcal{C}(\bbs)$, then
we have $(R_0,R_1,R_e-\gamma)\in\mathcal{C}_{\alpha}(\bbs)$. This
will let us write down an outer bound for $\mathcal{C}(\bbs)$ in
terms of $\mathcal{C}_\alpha(\bbs)$. To this end, we note that if
$(R_0,R_1,R_e)\in\mathcal{C}(\bbs)$, we need to have a random
vector $(U,V,\bbx)$ such that the inequalities given in
Theorem~\ref{theorem_csiszar} hold. Assume that we use the same
random vector $(U,V,\bbx)$ for the new Gaussian MIMO wiretap
channel in (\ref{channel_alpha_1})-(\ref{channel_alpha_2}), and
achieve the rate triple
$(\overline{R}_0,\overline{R}_1,\overline{R}_e)$. Due to the
Markov chains in (\ref{auxiliary_mc_1})-(\ref{auxiliary_mc_2}), we
already have $R_1\leq \overline{R}_1,R_0\leq \overline{R}_0$.
Furthermore, following the analysis in Section~4
of~\cite{Liu_Common_Confidential}, we can bound the gap between
$R_e$ and $\overline{R}_e$, i.e., $\gamma$, as follows
\begin{align}
\gamma=R_e-\overline{R}_e&\leq \frac{1}{2}
\log\frac{\left|\overline{\bbh}_Z\bbs\overline{\bbh}_Z^\top+\bbsigma_Z\right|}{|\bbsigma_Z|}-\frac{1}{2}\log\frac{|\bbh_Z\bbs
\bbh_Z^\top+\bbsigma_Z|}{|\bbsigma_Z|}
\end{align}
Thus, we have
\begin{align}
\mathcal{C}(\bbs)\subseteq
\mathcal{C}_\alpha(\bbs)+\mathcal{G}(\bbs)\label{dummy_outer_bound}
\end{align}
where $\mathcal{G}(\bbs)$ is
\begin{align}
\mathcal{G}(\bbs)=\left\{(0,0,R_e):0\leq R_e\leq\frac{1}{2}
\log\frac{\left|\overline{\bbh}_Z\bbs\overline{\bbh}_Z^\top+\bbsigma_Z\right|}{|\bbsigma_Z|}-\frac{1}{2}\log\frac{|\bbh_Z\bbs
\bbh_Z^\top+\bbsigma_Z |}{|\bbsigma_Z|}\right\}
\end{align}
Taking $\alpha\rightarrow 0$ in (\ref{dummy_outer_bound}), we get
\begin{align}
\mathcal{C}(\bbs)\subseteq \lim_{\alpha\rightarrow 0}
\mathcal{C}_\alpha(\bbs)\label{dummy_outer_bound_again}
\end{align}
where we use the fact that
\begin{align}
\lim_{\alpha \rightarrow 0} \frac{1}{2}
\log\frac{\left|\overline{\bbh}_Z\bbs\overline{\bbh}_Z^\top+\bbsigma_Z\right|}{|\bbsigma_Z|}-\frac{1}{2}\log\frac{|\bbh_Z\bbs
\bbh_Z^\top +\bbsigma_Z|}{|\bbsigma_Z|}=0
\end{align}
which follows from the continuity of $\log|\cdot|$ in positive
semi-definite matrices, and the fact that $\lim_{\alpha\rightarrow
0}\overline{\bbh}_Z=\bbh_Z$. Finally, we note that
\begin{align}
\lim_{\alpha\rightarrow 0}\mathcal{C}_\alpha (\bbs)
\end{align}
converges to the region given in Theorem~\ref{theorem_main_result}
due to the continuity of $\log|\cdot|$ in positive semi-definite
matrices and $\lim_{\alpha\rightarrow
0}\overline{\bbh}_Y=\bbh_Y,\lim_{\alpha\rightarrow
0}\overline{\bbh}_Z=\bbh_Z$; completing the proof.

\section{Conclusions}
We study the Gaussian MIMO wiretap channel in which a common
message is sent to both the legitimate user and the eavesdropper
in addition to the private message sent only to the legitimate
user. We first establish an equivalence between this original
definition of the wiretap channel and the wiretap channel with
public messages, in which the private message is divided into two
parts as the confidential message, which needs to be transmitted
in perfect secrecy, and public message, on which there is no
secrecy constraint. We next obtain capacity regions for both
cases. We show that it is sufficient to consider jointly Gaussian
auxiliary random variables and channel input to evaluate the
single-letter description of the capacity-equivocation region due
to~\cite{Korner}. We prove this by using channel
enhancement~\cite{Shamai_MIMO} and an extremal inequality
from~\cite{Liu_Compound}.

\appendices

\section{Proof of Lemma~\ref{lemma_equivalence}}
\label{proof_of_lemma_equivalence}

The proof of this lemma for $R_0=0$ is outlined
in~\cite[Problem~33-c]{Csiszar_book},~\cite{Ruoheng_Equivocation}.
We extend their proof to the general case of interest here. We
first note the inclusion $\mathcal{C}_p\subseteq \mathcal{C}$,
which follows from the fact that if
$(R_0,R_p,R_s)\in\mathcal{C}_p$, we can attain the rate triple
$(R_0,R_1=R_s+R_p,R_e=R_s)$, i.e.,
$(R_0,R_s+R_p,R_s)\in\mathcal{C}$. To show the reverse inclusion,
we use the achievability proof for Theorem~\ref{theorem_csiszar}
given in~\cite{Korner}. According to this achievable scheme, $W_1$
can be divided into two parts as $W_1=(W_p,W_s)$ with rates
$(R_1-R_e,R_e)$, respectively, and we have
\begin{align}
H(W_1|W_0,Z^n)&= H(W_p,W_s|Z^n,W_0)\\
&\geq H(W_s|Z^n,W_0)\\
&\geq H(W_s)-n\gamma_n
\end{align}
for some $\gamma_n$ which satisfies $\lim_{n\rightarrow
\infty}\gamma_n=0$. Hence, using this capacity achieving scheme
for $\mathcal{C}$, we can attain the rate triple
$(R_0,R_p=R_1-R_e,R_s=R_e)\in\mathcal{C}_p$. This implies
$\mathcal{C}\subseteq \mathcal{C}_p$; completing the proof of the
lemma.

\section{Proof of Lemma~\ref{lemma_KKT_conditions}}
\label{proof_of_lemma_KKT_conditions}

Since the program in
(\ref{optimization_Gaussian_again_1})-(\ref{optimization_Gaussian_again_2})
is not necessarily convex, the KKT conditions are necessary but
not sufficient. The Lagrangian for this optimization problem is
given by
\begin{align}
\mathcal{L}&= \mu_s R_s(\bbk)+\mu_p R_p+\lambda_Y
\left[R_p(\bbk)+R_{0Y}(\bbk)-R_p-R_0^*\right]+\lambda_Z
\left[R_p(\bbk)+R_{0Z}(\bbk)-R_p-R_0^*\right]\nonumber\\
&\quad + \beta_Y \left[R_{0Y}(\bbk)-R_0^*\right]+ \beta_Z
\left[R_{0Z}(\bbk)-R_0^*\right] + {\rm tr}(\bbk\bbm) +{\rm
tr}((\bbs-\bbk)\bbm_S)
\end{align}
where $\bbm,\bbm_S$ are positive semi-definite matrices, and
$\lambda_Y\geq 0, \lambda_Z\geq 0$, $\beta_Y\geq 0,\beta_Z\geq 0$.

The necessary KKT conditions that they need to satisfy are given
as follows
\begin{align}
\frac{\partial{\mathcal{L}}}{\partial R_p} \mid_{R_p=R_p^*}&=0 \label{proof_KKT_1}\\
\nabla_{\bbk}{\mathcal{L}}\mid_{\bbk=\bbk^*}&=\bzero \label{proof_KKT_2}\\
{\rm tr}(\bbk^*\bbm)&=0\label{proof_KKT_3}\\
{\rm tr}((\bbs-\bbk^*)\bbm_S)&=0\label{proof_KKT_4}\\
\lambda_Y \left[R_p(\bbk^*)+R_{0Y}(\bbk^*)-R_p^*-R_0^*\right]&=0\label{proof_KKT_5} \\
\lambda_Z
\left[R_p(\bbk^*)+R_{0Z}(\bbk^*)-R_p^*-R_0^*\right]&=0\label{proof_KKT_5_add} \\
\beta_Y (R_{0Y}(\bbk^*)-R_0^*)&=0\label{proof_KKT_6}\\
\beta_Z (R_{0Z}(\bbk^*)-R_0^*)&=0 \label{proof_KKT_6_add}
\end{align}
The first KKT condition in (\ref{proof_KKT_1}) implies
$\lambda_Y+\lambda_Z=\mu_p$. We define
$\lambda_Y=\mu_p\lambda,\lambda_Z=\mu_p\bar{\lambda}$ and
consequently, we have $0\leq \bar{\lambda}=1-\lambda\leq 1$. The
second KKT condition in (\ref{proof_KKT_2}) implies (\ref{KKT_1}).
Since ${\rm tr}(\bba\bbb)={\rm tr}(\bbb \bba)$ and ${\rm tr}(\bba
\bbb) \geq 0$ for $\bba\succeq \bzero,\bbb\succeq \bzero $,
(\ref{proof_KKT_3})-(\ref{proof_KKT_4}) imply
(\ref{KKT_2})-(\ref{KKT_3}). The KKT conditions in
(\ref{proof_KKT_5})-(\ref{proof_KKT_5_add}) imply (\ref{KKT_6}).
Furthermore, the KKT conditions in
(\ref{proof_KKT_5})-(\ref{proof_KKT_5_add}) state the conditions
that if $R_{0Y}(\bbk^*)>R_{0Z}(\bbk^*)$, $\lambda=0$, if
$R_{0Y}(\bbk^*)<R_{0Z}(\bbk^*)$, $\lambda=1$, and if
$R_{0Y}(\bbk^*)=R_{0Z}(\bbk^*)$, $\lambda$ is arbitrary, i.e.,
$0<\lambda<1$. Similarly, the KKT conditions in
(\ref{proof_KKT_6})-(\ref{proof_KKT_6_add}) imply (\ref{KKT_5}).

\section{Proof of Lemma~\ref{lemma_dummy_enhancement}}
\label{proof_of_dummy_enhancement}

We note the following identities
\begin{align}
(\mu_s-\mu_p\lambda)(\bbk^*+\tilde{\bbsigma}_Z)^{-1}&=(\mu_s-\mu_p\lambda)(\bbk^*+\bbsigma_Z)^{-1}+\bbm_S
\label{dummy_enhancement_again} \\
(\mu_s-\mu_p\lambda)(\bbk^*+\tilde{\bbsigma}_Z)^{-1}&=(\mu_s-\mu_p\lambda)(\bbk^*+\bbsigma_Y)^{-1}+\bbm
\label{dummy_enhancement_implies_1}
\end{align}
where (\ref{dummy_enhancement_again}) is due to
(\ref{dummy_enhancement}), and (\ref{dummy_enhancement_implies_1})
is obtained by plugging (\ref{dummy_enhancement_again}) into
(\ref{KKT_1_again}). Since $\bbm\succeq \bzero,\bbm_S\succeq
\bzero$,
(\ref{dummy_enhancement_again})-(\ref{dummy_enhancement_implies_1})
implies
\begin{align}
(\mu_s-\mu_p\lambda)(\bbk^*+\tilde{\bbsigma}_Z)^{-1}&\succeq(\mu_s-\mu_p\lambda)(\bbk^*+\bbsigma_Z)^{-1} \label{really_dummy_1}\\
(\mu_s-\mu_p\lambda)(\bbk^*+\tilde{\bbsigma}_Z)^{-1}&\succeq
(\mu_s-\mu_p\lambda)(\bbk^*+\bbsigma_Y)^{-1}
\label{really_dummy_2}
\end{align}
Using the fact that for $\bba\succ \bzero$, $\bbb\succ \bzero$, if
$\bba \preceq \bbb$, then $\bba^{-1}\succeq \bbb^{-1}$ in
(\ref{really_dummy_1})-(\ref{really_dummy_2}), we can get the
first and second parts of Lemma~\ref{lemma_dummy_enhancement}. We
next show the third part of Lemma~\ref{lemma_dummy_enhancement} as
follows
\begin{align}
(\bbk^*+\tilde{\bbsigma}_Z)^{-1}(\bbs+\tilde{\bbsigma}_Z)&=\bbi+(\bbk^*+\tilde{\bbsigma}_Z)^{-1}(\bbs-\bbk^*)\\
&=\bbi+\left[(\bbk^*+\bbsigma_Z)^{-1}+\frac{1}{\mu_s-\mu_p\lambda}\bbm_S\right](\bbs-\bbk^*)
\label{dummy_enhancement_implies_2}\\
&=\bbi+(\bbk^*+\bbsigma_Z)^{-1}(\bbs-\bbk^*)\label{KKT_3_implies}\\
&=(\bbk^*+\bbsigma_Z)^{-1}(\bbs+\bbsigma_Z)
\end{align}
where (\ref{dummy_enhancement_implies_2}) is due to
(\ref{dummy_enhancement_again}), and (\ref{KKT_3_implies}) comes
from (\ref{KKT_3}). The proof is complete.

\section{Proof of Lemma~\ref{lemma_one_more_time}}

\label{proof_of_lemma_one_more_time}

We note the following
\begin{align}
(\mu_s-\mu_p\lambda)(\bbk^*+\tilde{\bbsigma}_Y)^{-1}&=(\mu_s-\mu_p\lambda)(\bbk^*+\bbsigma_Y)^{-1}+\bbm\label{one_more_time_again}\\
(\mu_s-\mu_p\lambda)(\bbk^*+\tilde{\bbsigma}_Y)^{-1}&=(\mu_s-\mu_p\lambda+\mu_0\bar{\beta})(\bbk^*+\bbsigma_Z)^{-1}
+\mu_0\beta
(\bbk^*+\bbsigma_Y)^{-1}+\bbm_S\label{one_more_time_implies}
\end{align}
where (\ref{one_more_time_again}) is (\ref{one_more_time}), and
(\ref{one_more_time_implies}) comes from plugging
(\ref{one_more_time_again}) into (\ref{KKT_1_again_1}). Since
$\bbm\succeq \bzero$, (\ref{one_more_time_again}) implies
\begin{align}
(\mu_s-\mu_p\lambda)(\bbk^*+\tilde{\bbsigma}_Y)^{-1}&\succeq(\mu_s-\mu_p\lambda)(\bbk^*+\bbsigma_Y)^{-1}
\label{one_more_time_again_implies}
\end{align}
Using the fact that for $\bba\succ \bzero$, $\bbb\succ \bzero$, if
$\bba \preceq \bbb$, then $\bba^{-1}\succeq \bbb^{-1}$ in
(\ref{one_more_time_again_implies}) yields the first statement of
the lemma. Since $0\leq \beta=1-\bar{\beta}\leq 1$ and
$\bbm_S\succeq \bzero$, (\ref{one_more_time_implies}) implies
\begin{align}
(\mu_s-\mu_p\lambda)(\bbk^*+\tilde{\bbsigma}_Y)^{-1}&\succeq
(\mu_s-\mu_p\lambda)(\bbk^*+\bbsigma_Z)^{-1}
\label{one_more_time_implies_implies}
\end{align}
Using the fact that for $\bba\succ \bzero$, $\bbb\succ \bzero$, if
$\bba \preceq \bbb$, then $\bba^{-1}\succeq \bbb^{-1}$ in
(\ref{one_more_time_implies_implies}) yields the second statement
of the lemma. We next consider the third statement of this lemma
as follows
\begin{align}
(\bbk^*+\tilde{\bbsigma}_Y)^{-1}\tilde{\bbsigma}_Y&=\bbi-(\bbk^*+\tilde{\bbsigma}_Y)^{-1}\bbk^*\\
&=\bbi-\left[(\bbk^*+\bbsigma_Y)^{-1}+\frac{1}{\mu_s-\mu_p\lambda}\bbm\right]\bbk^*
\label{one_more_time_again_implies_again}\\
&=\bbi-(\bbk^*+\bbsigma_Y)^{-1}\bbk^*
\label{KKT_2_implies} \\
&=(\bbk^*+\bbsigma_Y)^{-1}\bbsigma_Y
\end{align}
where (\ref{one_more_time_again_implies_again}) is due to
(\ref{one_more_time_again}) and (\ref{KKT_2_implies}) comes from
(\ref{KKT_2}).

\section{Proof of Lemma~\ref{lemma_explicit_form}}

\label{proof_of_lemma_explicit_form} The optimization problem in
(\ref{optimization_for_contradiction_again_1}) can be written as
\begin{align}
&\max_{\substack{U\rightarrow V\rightarrow \bbx \rightarrow
(\bby,\bbz)\\E\left[\bbx \bbx^\top\right] \preceq \bbs}}~~\mu_s
R_s+\mu_p
R_p +(\mu_p+\beta_Z)R_0 \label{cost_function}\\
&\qquad~~ {\rm s.t.} \left\{
\begin{array}{rcl}
0\leq R_s &\leq &I(V;\bby|U)-I(V;\bbz|U) \\
R_s+R_p+R_0 &\leq & I(V;\bby|U)+\min\{I(U;\bby),I(U;\bbz)\}\\
R_0 &\leq & \min\{I(U;\bby),I(U;\bbz)\}
\end{array}\right.
\label{constraint_set}
\end{align}
For a given $(U,V,\bbx)$, we can rewrite the cost function in
(\ref{cost_function}) as follows
\begin{align}
\lefteqn{\mu_sR_s+\mu_pR_p+(\mu_p+\beta_Z)R_0}\nonumber\\
&\leq \mu_s R_s+\mu_p
[I(V;\bby|U)+\min\{I(U;\bby),I(U;\bbz)\}-R_s-R_0]
+(\mu_p+\beta_Z)R_0 \label{second_constraint}\\
&=(\mu_s-\mu_p)R_s+\mu_p
[I(V;\bby|U)+\min\{I(U;\bby),I(U;\bbz)\}]+\beta_Z R_0 \\
&\leq (\mu_s-\mu_p)[I(V;\bby|U)-I(V;\bbz|U)]+\mu_p
[I(V;\bby|U)+\min\{I(U;\bby),I(U;\bbz)\}]+\beta_Z R_0
\label{first_constraint} \\
&=\mu_s[I(V;\bby|U)-I(V;\bbz|U)]+\mu_p
[I(V;\bbz|U)+\min\{I(U;\bby),I(U;\bbz)\}]+\beta_Z R_0\\
&\leq \mu_s[I(V;\bby|U)-I(V;\bbz|U)]+\mu_p
[I(V;\bbz|U)+\min\{I(U;\bby),I(U;\bbz)\}]\nonumber\\
&\quad +\beta_Z\min\{I(U;\bby),I(U;\bbz)\}\label{third_constraint}\\
&= \mu_s[I(V;\bby|U)-I(V;\bbz|U)]+\mu_p I(V;\bbz|U)
+(\mu_p+\beta_Z)\min\{I(U;\bby),I(U;\bbz)\}\label{most_dummiest}
\end{align}
where (\ref{second_constraint}) comes from the second constraint
in (\ref{constraint_set}), (\ref{first_constraint}) is due to the
first constraint in (\ref{constraint_set}) and the assumption
$\mu_s>\mu_p$, and (\ref{third_constraint}) comes from the third
constraint in (\ref{constraint_set}). The proof can be concluded
by noting that the upper bound on the cost function given in
(\ref{most_dummiest}) is attainable.

\bibliographystyle{unsrt}
\bibliography{IEEEabrv,references2}
\end{document}